\documentclass[a4paper]{article}

\usepackage{INTERSPEECH2022}
\usepackage{amsmath,graphicx}
\usepackage{amssymb}
\usepackage{hyperref}
\usepackage{multirow}
\usepackage{booktabs}
\usepackage[table, svgnames, dvipsnames]{xcolor}
\usepackage[normalem]{ulem}
\usepackage{todonotes} 
\usepackage{float}
\makeatletter
\newcommand\footnoteref[1]{\protected@xdef\@thefnmark{\ref{#1}}\@footnotemark}
\makeatother
\definecolor{lightergray}{gray}{0.85}

\definecolor{mygreen}{HTML}{008000}

\newcolumntype{a}{>{\columncolor{lightergray}}c}
\newcolumntype{b}{>{\columncolor{lightergray}}l}

\title{Word Discovery in Visually Grounded, Self-Supervised Speech Models}
\name{Puyuan Peng, David Harwath}
\address{Department of Computer Science, The University of Texas at Austin}
\email{pyp@utexas.edu, harwath@utexas.edu \\ 
\vspace{3mm}
\includegraphics[width=1\textwidth]{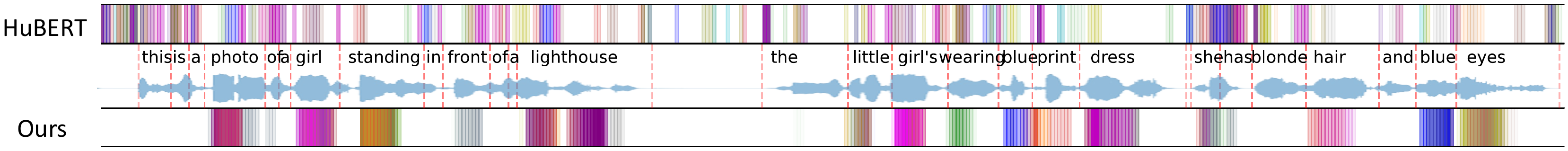} 
\vspace{-6mm}
\captionof{figure}{HuBERT: sum of attention weights each frame receives from other frames. Ours ($\text{VG-HuBERT}_3$): attention weights each frame receives from the \texttt{[CLS\_A]} token. Attention weights from different attention heads are coded with different colors.}\label{fig:girl_lighthouse}
\vspace{-5mm}}

\begin{document}

\maketitle
\begin{abstract}
We present a method for visually-grounded spoken term discovery. After training either a HuBERT or wav2vec2.0 model to associate spoken captions with natural images, we show that powerful word segmentation and clustering capability emerges within the model's self-attention heads. Our experiments reveal that this ability is not present to nearly the same extent in the base HuBERT and wav2vec2.0 models, suggesting that the visual grounding task is a crucial component of the word discovery capability we observe. We also evaluate our method on the Buckeye word segmentation and ZeroSpeech spoken term discovery tasks, where we perform on par with or better than currently published methods on several metrics.
\end{abstract}

\vspace{-1mm}
\section{Introduction and Related Work}
\vspace{-1mm}
The task of spoken term discovery involves starting with a corpus of untranscribed, unsegmented speech audio data and developing an algorithm or computational model that is capable of 1) inferring the temporal location of each word boundary and 2) clustering the resulting speech segments into categories that reflect word identity. The motivation behind this task is twofold. First, a model capable of solving this problem would be immensely useful for building speech understanding systems using a limited amount of transcribed data that could be applied to low-resource languages and domains. Second, the model could be used by cognitive science researchers as a model for the process by which human children acquire spoken language~\cite{dupoux18}.

The spoken term discovery task has a long history dating back at least to the seminal work of Park and Glass~\cite{park08}, who proposed a segmental dynamic time warping algorithm to solve both the segmentation and clustering sub-tasks. In the time since then, numerous other approaches have been proposed, based on techniques such as refinements to the S-DTW algorithm~\cite{Jansen2010TowardsST}, Bayesian probabilistic models~\cite{lee15, besgmm}, dynamic programming algorithms~\cite{kamper17,Kamper2021TowardsUP,Kamper2022WordSO} and contrastive predictive coding with neural models~\cite{vandenoord18,Cuervo2021ContrastivePS,bhati21}. The ZeroSpeech~\cite{Versteegh2017TheZR} challenge was also introduced to provide a standard benchmark to the community for the word discovery task, in addition to other related tasks in zero-resource speech processing.

A separate line of research has investigated training speech models using a visual grounding (VG) objective~\cite{synnaeve14, harwath15, Harwath2016UnsupervisedLO, chrupala21}. Given untranscribed, unsegmented speech that semantically describes an auxiliary input from the visual modality, a model is trained to associate the inputs together. Numerous works have demonstrated that these models' learned representations capture linguistic abstractions such as phonemes and words~\cite{Harwath2018, Havard2019WordRC, Harwath2020LearningHD, Khorrami2021CanPS, nikolaus22, Scholten2021LearningTR, Olaleye21, 9723367, Liu2021CrossModalDR}, but so far none of them have explicitly used a VG model on the spoken term discovery task.

In this paper, we present a method for \textit{visually-grounded} spoken term discovery\footnote{code and model weights will be made available at \href{https://github.com/jasonppy/word-discovery}{https://github.com/jasonppy/word-discovery}} using a Transformer~\cite{vaswani17} model. Our method is simple: it simply involves applying a binary threshold to the self-attention maps produced by the model, and extracting contiguous temporal regions of the speech signal with attention scores above the threshold. We find that these segments accurately predict both the boundaries and identity of the underlying words, which we verify with a suite of experiments on the SpokenCOCO dataset. Additionally, we evaluate our proposed method on the Buckeye word segmentation~\cite{buckeye} and ZeroSpeech~\cite{Versteegh2017TheZR} English spoken term discovery tasks, and find in both cases it significantly outperforms the current state-of-the-art, non-visually-grounded models from the literature. 

\vspace{-2mm}
\section{Models}
\vspace{-2mm}
Our model has a simple Transformer-based dual-encoder architecture consisting of an audio encoder and an image encoder. The model takes as input a raw speech waveform $A$ and an image $I$, and outputs a similarity score that should be large when the speech faithfully describes the content of the image, and small otherwise. 
The audio encoder has the same architecture as the wav2vec2.0 (W2V2) and HuBERT Base models. The input speech utterance is represented as waveform samples which are processed by a convolutional block to obtain features of a higher temporal resolution. Convolutional positional embeddings are added to the output features. We then follow \cite{peng2021} and augment the features with a \texttt{[CLS\_A]} token before feeding them to the stack of Transformer layers. The image encoder is a Vision Transformer (ViT)~\cite{vit,dino}. It takes an RGB image and splits it into a grid of non-overlapping patches, then uses a linear layer to embed the patches into a sequence of feature vectors. Trainable positional embeddings are added to the features, which are then concatenated with a \texttt{[CLS\_I]} token before being fed to the Transformer layers. The similarity score between image $I$ and audio $A$ is defined as the dot product of \texttt{[CLS\_I]} and \texttt{[CLS\_A]}. Given parallel speech-image data, we can train the model using the InfoNCE loss\cite{vandenoord18,Ilharco2019LargeScaleRL}, where positive samples are matched pairs and negative samples unmatched pairs. The loss encourages high similarity scores between matched image-speech pairs and low similarity scores otherwise.

Unlike most models used in vision and language research, none of the modules in our model are pretrained on any supervised tasks such as image classification, object detection or automatic speech recognition. Both our image and audio encoders are pretrained in a completely self-supervised fashion. We will refer to our visually-grounded variants of HuBERT and W2V2 as \textbf{VG-HuBERT} and \textbf{VG-W2V2}. 
As will be shown in section~\ref{sec:exp}, similar to~\cite{Pasad2021LayerWiseAO}, we found that reinitializing the last few Transformer layers of the pretrained HuBERT or W2V2 with random weights can significantly improve the word discovery performance. We name models that reinitialize weights of the last $x$ Transformer layers \textbf{$\text{VG-HuBERT}_{x}$} and  \textbf{$\text{VG-W2V2}_{x}$}. 
\vspace{-2mm}
\section{Experiments}\label{sec:exp}
\vspace{-2mm}
\textbf{Datasets.}
We train our models on the visual grounding task with the Karpathy training split~\cite{Karpathy2017DeepVA} of the SpokenCOCO dataset~\cite{Hsu2020TextFreeIS}. SpokenCOCO contains $123$k images, each with $5$ spoken captions produced by humans speaking aloud the text captions in MSCOCO~\cite{Lin2014MicrosoftCC}. We evaluate our model using both speech-image retrieval and word discovery tasks on the SpokenCOCO Karpathy test set. We additionally evaluate our model on the spoken term discovery task using a 6 hour test set~\cite{Kamper2022WordSO} of the Buckeye corpus~\cite{buckeye}, as well as the ZeroSpeech 2020~\cite{Versteegh2017TheZR} spoken term discovery evaluation split.

\textbf{Implementation Details.}
Our image encoder inherits the architecture and is initialized from the DINO~\cite{dino} self-supervised Vision Transformer. In particular, we initialize the image encoder weights from DINO-ViT small with patch size 8x8. The model consists of a linear embedding layer, a learnable positional embedding, and a $12$-layer transformer with hidden dimension of $384$ and $6$ attention heads. The audio encoder inherits its architecture and initial weights from HuBERT (or W2V2) Base, which contains a $7$-layer convolutional block, and a $12$-layer Transformer with a hidden dimension of $786$ and $12$ attention heads. We project the output \texttt{[CLS\_I]} and \texttt{[CLS\_A]} separately using $2$-layer MLPs to a $2048$ dimensional space before taking their dot product. Models are trained on the visual grounding task with batch size $100$ for $30$ epochs. We optimize with BertAdam~\cite{Devlin2019BERTPO} using a learning rate linearly ramped up $5e-5$ over the first 10\% of training, and then linearly decayed to $0$. Except for the convolutional block in the audio encoder that is frozen, all other layers are trainable.

\textbf{Retrieval Performance.} As an initial sanity check, we first quantify how well our models learn to associate spoken descriptions with images by evaluating them on an image-speech retrieval task. We report Recall@K for $K=1, 5,10$ in Table~\ref{tab:retrival}. Although VG-HuBERT and $\text{VG-HuBERT}_3$ do not use any supervised vision modules, they still perform relatively well on this benchmark compared to $\text{FaST-VGS}_{\text{CO}}$, which is the current SoTA dual-encoder retrieval model that uses a pretrained Faster RCNN for image feature extraction.
\begin{table}[htb]
\vspace{-3mm}
  \caption{Comparison of VG-HuBERT with other models on the 1k retrieval task on the SpokenCOCO test set.}
  \vspace{-6mm}
  \begin{center}
  \resizebox{\columnwidth}{!}{
  \begin{tabular}{lcccccc}
      \toprule
      &\multicolumn{3}{c}{Speech $\rightarrow$ Image}&\multicolumn{3}{c}{Image $\rightarrow$ Speech}\\\cmidrule(lr){2-4} \cmidrule(lr){5-7}  
      Model&R@1&R@5&R@10 &R@1&R@5&R@10\\
      \midrule
      ResDAVEnet~\cite{Hsu2019TransferLF} &17.3 &41.9 &55.0 & 22.0&50.6 &65.2\\
      $\text{FaST-VGS}_{\text{CO}}$~\cite{peng2021} & \underline{31.8}& \underline{62.5}& \underline{75.0}& 42.5 &\underline{73.7} &\underline{84.9} \\
      $\text{FaST-VGS}_{\text{CTF}}$~\cite{peng2021} & \textbf{35.9}& \textbf{66.3}& \textbf{77.9}& \textbf{48.8} &\textbf{78.2} &\textbf{87.0} \\
      VG-HuBERT &30.4&60.7&72.7&42.7&73.6&84.1 \\
      $\text{VG-HuBERT}_{3}$ &30.6&60.8&72.8&\underline{42.8}&73.5&83.9 \\
      \bottomrule
  \end{tabular}}
  \end{center}
  \label{tab:retrival}
  \vspace{-6mm}
  \end{table}

\begin{table*}[ht]
\caption{Different models evaluated on their word discovery ability on the on SpokenCOCO test. $\pm$ indicates the standard deviation across 5 clustering runs with different random seeds; all models used 4096 k-means clusters. HuBERT$_{FT}$ and W2V2$_{FT}$ indicate models that were further fine-tuned on the SpokenCOCO audio captions.}\label{tab:spokencoco_main_result}
\vspace{-7mm}
    \begin{center}
    \resizebox{\textwidth}{!}{%
    \begin{tabular}{lcccaccacblb}
    \toprule
        &\multicolumn{4}{c}{Area} & \multicolumn{5}{c}{Boundary} & \multicolumn{2}{c}{Word} \\ \cmidrule(lr){2-5} \cmidrule(lr){6-10} \cmidrule(lr){11-12}
        Model & WC & tIoU & CD (ms) & A-score& Precision & Recall & $F_1$ & OS & $R$-val  &Purity& WD \\ 
      \cmidrule(lr){1-12}
      ResDAVEnet-VQ~\cite{Harwath2020Learning} & 98.39 & 23.78  & 119.4&38.30 & 10.42 & 50.96 &  17.30& 38.88 &  -250.77 &31.6 $\pm$ 0& 262 $\pm$ 0 \\
      W2V2~\cite{baevski20}&77.68&38.56&111.9&51.54&11.52&24.33&15.63&11.12&-33.34 & 41.2 $\pm$ 0.3 &407 $\pm$ 13 \\
      $\text{W2V2}_\text{FT}$~\cite{baevski20} & 73.96&39.87&116.1&51.81&11.88&24.79&16.06&10.87&-31.10&43.0 $\pm$ 0.1 &397 $\pm$ 11 \\
      HuBERT~\cite{hsu2021hubert}&72.63&37.72&117.9&49.66&12.18&24.97&16.37&10.51&-28.26 & 43.1 $\pm$ 0.2 &404 $\pm$ 12 \\
        $\text{HuBERT}_\text{FT}$~\cite{hsu2021hubert} &73.78&39.31&116.0&51.30&11.90&25.81&16.29&11.68&-36.72&45.3 $\pm$ 0.3 &500 $\pm$ 9 \\
        FaST-VGS~\cite{peng2021}   & 75.92 & 54.86   & 67.8 & 63.69& 28.99 & 26.17 & \underline{27.51} & -9.72 & \underline{40.10} &70.9 $\pm$ 0.5  & 1011 $\pm$ 12\\
        FaST-VGS+~\cite{peng2022}  & 77.40&51.46&80.5&61.82& 22.66 & 27.86 & 24.99 & 22.93 & 28.54 & 72.3 $\pm$ 0.3 & 1026 $\pm$ 12 \\ 
        VG-W2V2 &65.94&45.56&76.8&53.89&18.47&19.78&19.10&7.09&28.86& 62.5 $\pm$ 0.4 &743 $\pm$  14 \\
        $\text{VG-W2V2}_4$ &71.28&54.06 &82.0& 61.49&28.15&22.90&25.26&-18.64&39.67& 73.0 $\pm$ 0.3 &\underline{1182} $\pm$ 20 \\
        $\text{VG-W2V2}_5$ &73.32&53.09 &72.2&61.58&28.70&25.45&26.98&-11.32&39.94 & 75.4 $\pm$ 0.1 &1149 $\pm$ 16\\
        VG-HuBERT  & 73.16 & 44.02   & 89.0& 54.97 & 18.31 & 18.90 & 18.60 & 3.26 & 29.60 &63.2 $\pm$ 0.4 & 823 $\pm$ 20 \\
        $\text{VG-HuBERT}_3$  &70.94&61.29&64.6&\textbf{65.77}& 35.90 & 27.03 & \textbf{30.84} & -24.72 & \textbf{44.42} & 75.3 $\pm$ 0.2 & 1167 $\pm$ 26 \\ 
        $\text{VG-HuBERT}_4$  &73.97 & 56.35   & 78.9& \underline{63.97} & 28.39 & 25.64 & 26.94 & -9.70 & 39.64 & 75.2 $\pm$ 0.2 & \textbf{1230} $\pm$ 18 \\
        \bottomrule
    \end{tabular}}
    \end{center}
    \vspace{-8mm}
\end{table*}

\begin{figure}
  \centering
      \includegraphics[width=1\columnwidth]{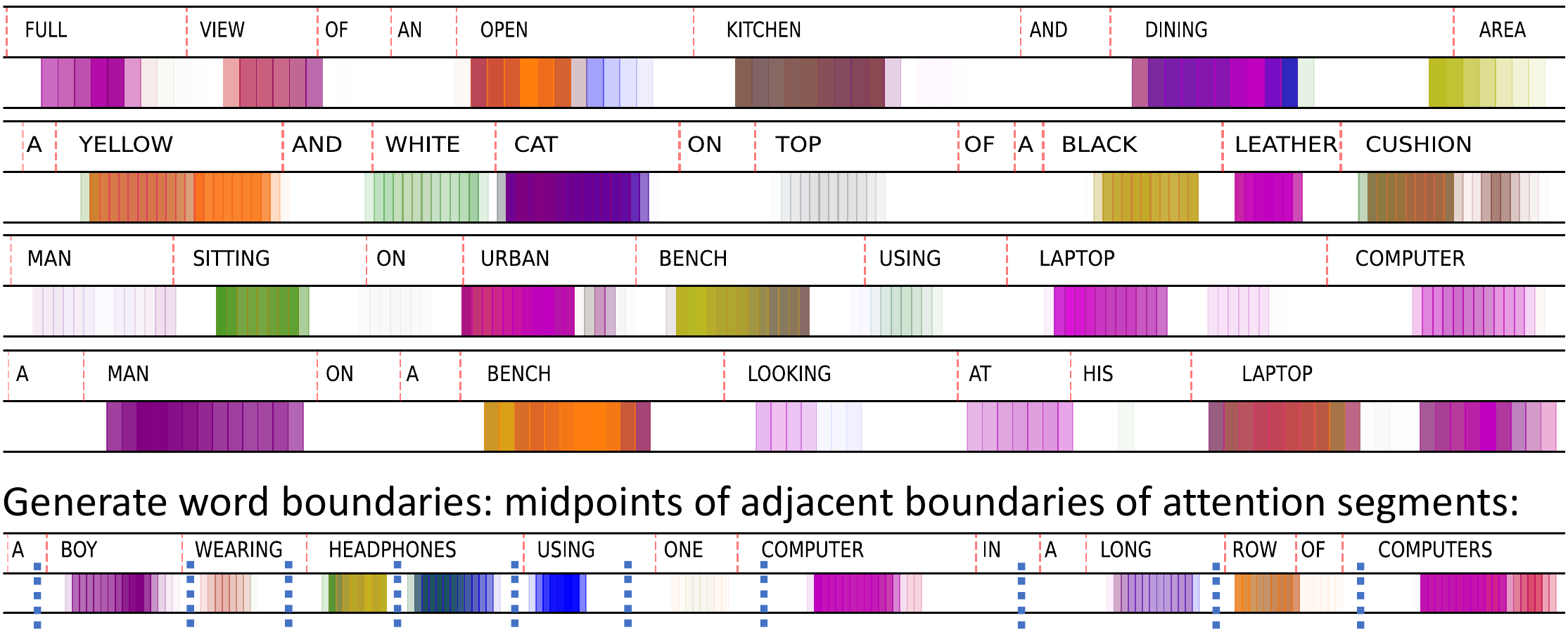} 
      \vspace{-6mm}
      \caption{Examples of CLS attention segments from SpokenCOCO, the bottom one also show how word boundaries are generated from attention segments}\label{fig:more_eg}
      \vspace{-5mm}
\end{figure}

\textbf{Word Discovery.} The main focus of this paper is studying the emergence of word segmentation in visually grounded speech models (the image branch is not used in any word discovery experiment). We first discovered this phenomenon by plotting the \texttt{[CLS\_A]} token's attention weight across each temporal frame of the speech at layer $y$ of the audio encoder;  Figure~\ref{fig:more_eg} displays several examples using $y=12$ from the $\text{VG-HuBERT}_{3}$ model. We threshold the weights to retain only the top $z\%$ of the total attention weight for each attention head ($z=10$).  The ground truth transcripts and force aligned word boundaries (red dashed lines) are also plotted for reference. Figure~\ref{fig:girl_lighthouse} is generated the same way with $y=9, z=10$.
Although we do not impose any constraints on the attention weights except for the thresholding described above, the visualization shows that the our model exhibits word localization, segmentation, and identification as it puts chunks of attention associated with different heads (indicated by different colors) under different words. In the following paragraphs, we pose several questions and metrics to quantitatively investigate these phenomena.

\textbf{1. Does the model's attention map indicate the presence of words?} 
We define two metrics to study this question:
\textbf{word coverage (WC)}, defined as $\frac{\#\text{hits}}{\#\text{words}}$, where $\#\text{hits}$ is the number of times the model assigns one or more attention segments to a word, and $\#\text{words}$ is the count of total words in the corpus. We define an attention segment to be a consecutive sequence of frames with nonzero attention weights (after thresholding has been applied), and consider a segment to be assigned to a particular word if more than half of the segment's length falls between the ground-truth boundaries of that word.
We also define \textbf{Temporal intersection over union (tIoU)} as $\frac{1}{|\mathcal{S}|} \sum_{s\in \mathcal{S}} \frac{s \cap w_s}{s \cup w_s}$, where $\mathcal{S}$ is the set of all attention segments, $w_s$ is the word that attention segment $s$ is assigned to, and the output of the intersection (or union) operation is the duration in frames of the intersection (or union) between the segment spanned by $w_s$ and the segment spanned by $s$. WC and tIoU are complementary to each other --- WC reflects how good the model is at detecting all words within an utterance, and tIoU shows how good the model is at localizing the span of each word it detects. A model could achieve a high WC if it assigns an attention segment every few frames (similar to HuBERT's attention map shown in Figure~\ref{fig:girl_lighthouse}), but this will lead to a low tIoU; on the other hand, a model might achieve a good tIoU score if it ignores most words, and places attention segments only under the words that it is very confident about (e.g. salient visual words). To balance these metrics, we introduce a summary metric named \textbf{area score (A-score)}, defined as the harmonic mean of WC and tIoU, i.e. $\text{A-score} = \frac{2\text{WC}\cdot \text{tIoU}}{\text{WC} + \text{tIoU}}$. In addition, we also show \textbf{center distance (CD)} which is defined as the average distance in milliseconds between the center of each attention segment and center of the word that the segment is assigned to. We group the all above metrics under the name ``Area''.

\begin{figure*}[h]
  \centering
  \begin{minipage}{0.75\columnwidth}
\captionof{table}{Performance of $\text{VG-HuBERT}_x$ on SpokenCOCO validation set.}\label{tab:diff_reinit}
\vspace{-6mm}
    \begin{center}
    \resizebox{\columnwidth}{!}{%
    \begin{tabular}{lccccc}
    \toprule
    &\multicolumn{3}{c}{Area} & \multicolumn{1}{c}{Boundary} & \multicolumn{1}{c}{Word} \\ \cmidrule(lr){2-4} \cmidrule(lr){5-5} \cmidrule(lr){6-6}
reinit last x  & WC & tIoU & A-score & $F_1$ & WD \\
\midrule
        x=0  &73.21&44.16 & 55.09 & 18.46  & 817 \\
        x=1  &69.29&51.88& 59.33 & 22.17 & 1003 \\
        x=2  &69.63&57.73& 63.12 &\underline{28.52} & 1036 \\
        x=3   &70.95&61.51& \textbf{65.91}& \textbf{31.07} & \underline{1169} \\
        x=4  & 73.90& 56.32& \underline{63.92}& 27.02 & \textbf{1229} \\
        x=5  &74.92&54.44 & 63.05& 25.52 & 1044 \\
        x=6 &73.43&51.38  & 60.45& 23.30 & 879 \\
\bottomrule
    \end{tabular}}
    \end{center}
  \end{minipage}\hfill
  \begin{minipage}{0.3\textwidth}
      \centering
      \includegraphics[width=1.1\columnwidth]{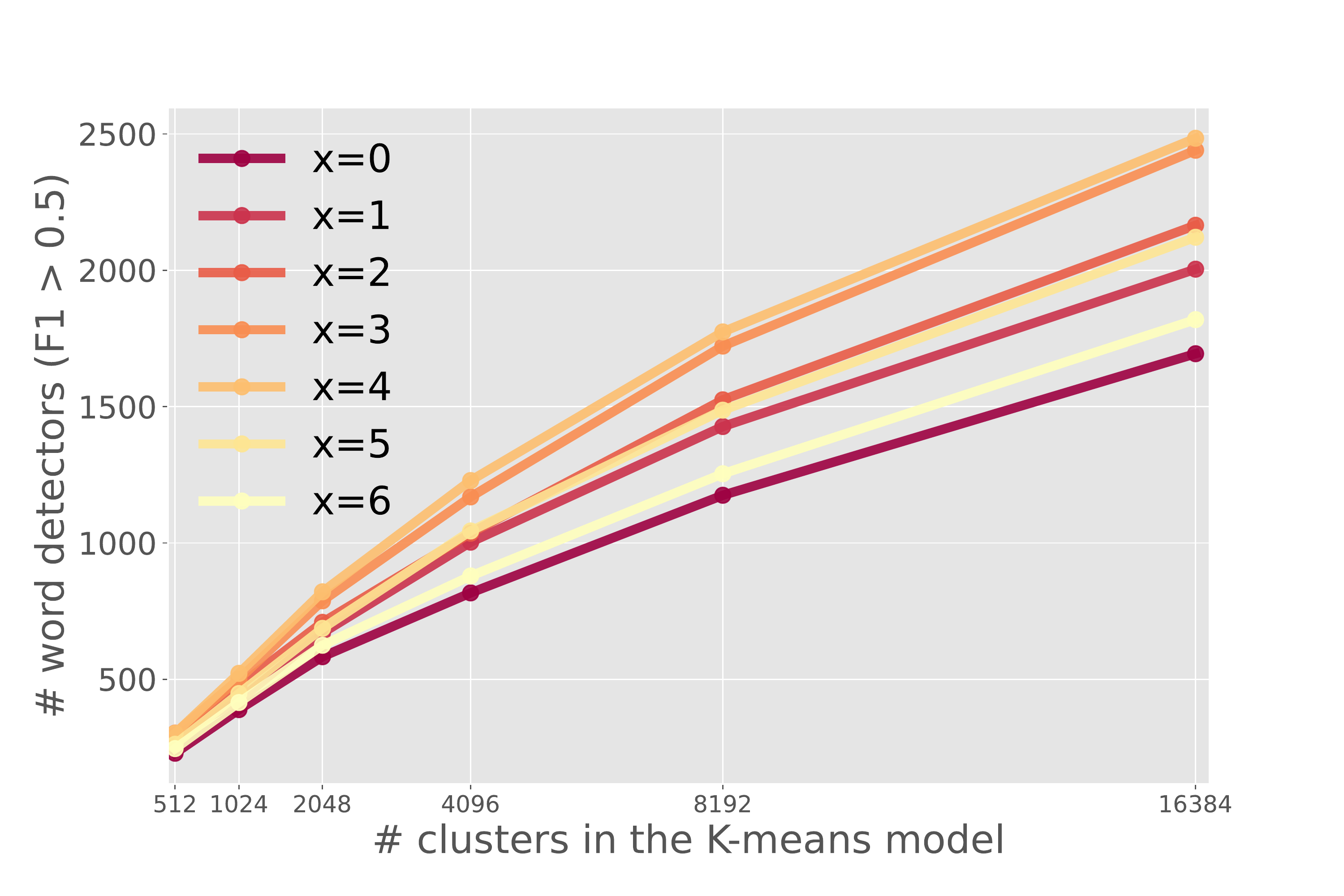} 
      \vspace{-8mm}
      \caption{\footnotesize \#WD given by $\text{VG-HuBERT}_x$ with varied \#K-means clusters.}\label{fig:diff_k}
  \end{minipage}\hfill
  \begin{minipage}{0.3\textwidth}
      \centering
      \includegraphics[width=1.13\columnwidth]{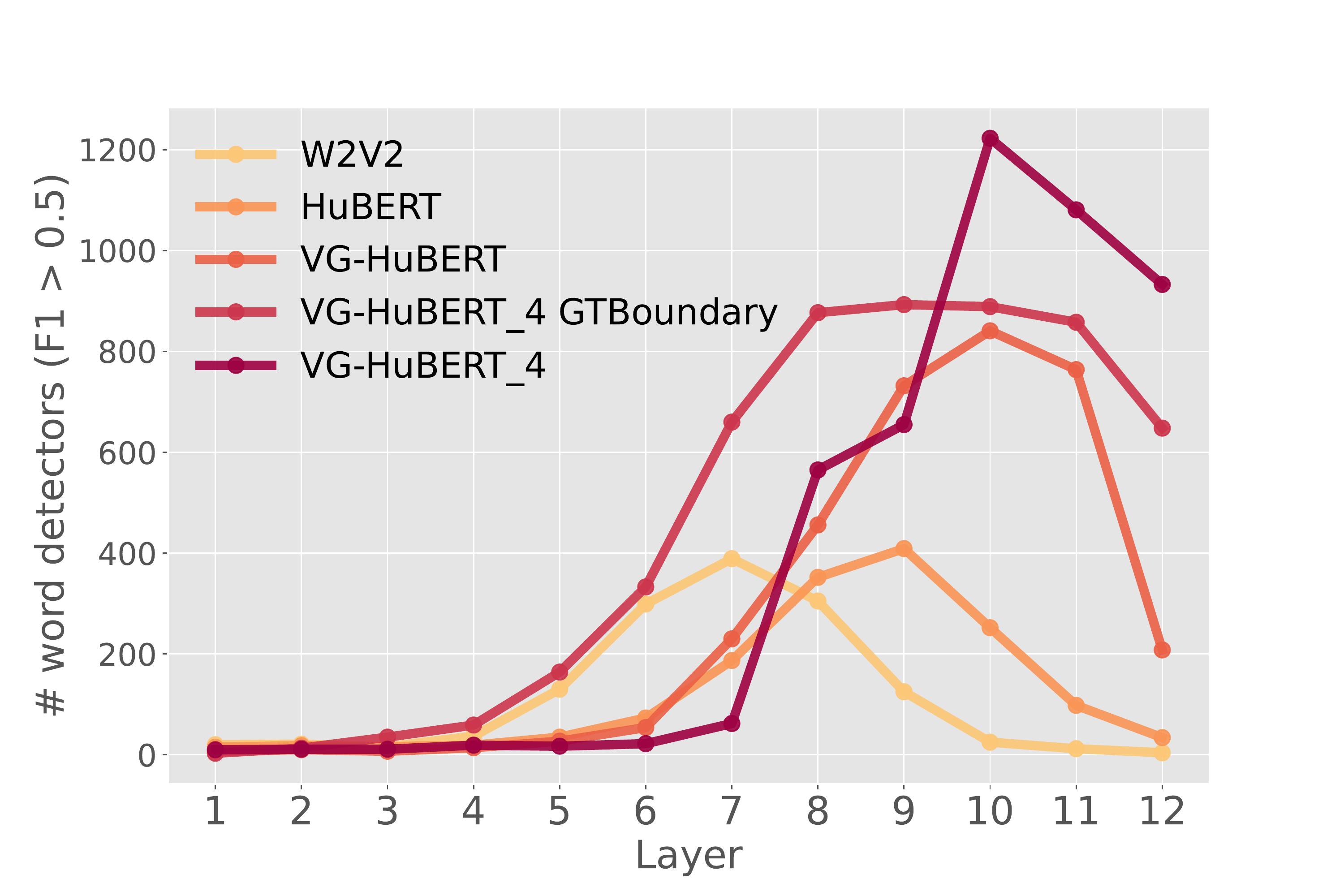}
      \vspace{-8mm}
      \caption{\footnotesize \#WD given by different models across all layers.}\label{fig:diff_layers}
  \end{minipage}
  \vspace{-6mm}
\end{figure*}
\textbf{2. Can we use the model to identify word boundaries?} Here, we are interested in seeing whether we can use the model to precisely predict the onset and offset of each word in an utterance. For evaluation, we use the standard boundary \textbf{precision}, \textbf{recall}, $\bm{F_1}$, \textbf{over-segmentation (OS)}, and \textbf{$\bm{R}$-value} metrics~\cite{Kamper2022WordSO}. These metrics are quite strict and only assume a very small tolerance window (20ms for SpokenCOCO and Buckeye~\cite{Kamper2022WordSO} and 30ms for ZeroSpeech~\cite{Versteegh2017TheZR}). However, we note that visually, the boundaries of our thresholded attention segments are typically narrower than the words they are assigned to. To address this, we instead infer word boundaries to be the midpoints between the edges of adjacent attention segments (example shown at the bottom in Figure~\ref{fig:more_eg}). We group the metrics used for studying question 2 under the name ``Boundary''. 

\textbf{3. Can the model determine word identity?} Here, we use the word detection evaluation introduced by~\cite{Harwath2020Learning}. To do so, we first average pool the representations belonging to each attention segment to obtain a single feature vector representing that segment. We apply this procedure to all utterances in the SpokenCOCO test set, and then use a K-means clustering algorithm to group all of the segments across the test set. Next, following~\cite{Harwath2020Learning}, we consider each cluster as a potential ``detector'' for every different vocabulary word appearing in the ground-truth transcripts. We compute precision, recall, and F1 scores for each unique cluster-word pair, and consider a cluster to be a ``word detector'' if it achieves an F1 score greater than or equal to 0.5 for some vocabulary word. Finally, we count the number of word-detecting clusters and report that as the \textbf{word detection (WD)} score. We additionally report cluster \textbf{purity} (in terms of the identity of each attention segment's assigned word) as a way of measuring the clustering performance. We group the metrics used for studying word detection under the name ``Word''.

\textbf{Hyperparameter tuning.} Performance on all of the above metrics depends on the model layer being examined, as well as the threshold applied to the attention maps. We tune these hyperparameters on the SpokenCOCO dev set separately for each metric group (Area, Boundary, and Word), respectively optimizing for A-score, boundary F1, and the number of word detectors found. During this tuning, we consider all 12 Transformer layers. For the attention threshold we sweep the threshold between 0.5 and 0.99. After tuning, we report all final results on the SpokenCOCO test set.

\textbf{Results on SpokenCOCO.} We summarize our results across the Area, Boundary, and Word metrics on the SpokenCOCO test dataset in Table~\ref{tab:spokencoco_main_result}. We compare against several visually-grounded baseline models from the literature, including ResDAVEnet-VQ~\cite{Harwath2020LearningHD}, FaST-VGS~\cite{peng2021} and FaST-VGS+~\cite{peng2022}. For ResDAVEnet-VQ, we use the layer 3 representations from the $\{3\}\rightarrow\{2,3\}$ model from~\cite{Harwath2020LearningHD} distributed by the authors as it achieved the highest word detection performance in \cite{Harwath2020LearningHD}. We first fine-tune ResDAVEnet-VQ on the SpokenCOCO data and then derive segments from consecutive frames assigned to the same VQ3 codebook entry. For the FaST-VGS models, we consider all layers of the \texttt{trm1} module and report results for layer achieving the best results on the dev set. Because \texttt{trm1} is also a transformer stack augmented with a \texttt{[CLS\_A]} token, we derive attention segments from these models in the same manner as for VG-HuBERT. Finally, we compare against the Librispeech-pretrained HuBERT and wav2vec2.0 Base models, and these models fine-tuned on audio of SpokenCOCO. 
Although these are non-visually grounded models, they share the same architecture as VG-HuBERT and VG-W2V2 so making this comparison allows us to probe whether the same architecture trained solely with a masked language modeling objective will still exhibit emergent word discovery ability in its self-attention maps. 
Because W2V2 and HuBERT don't use a \texttt{[CLS\_A]} token, we use the sum of attention weights each frame receives from other frames (separate summation for each attention head) to get the self-attention map, and then segments are derived in the same manner as for VG-HuBERT.

In Table~\ref{tab:spokencoco_main_result}, we first note that the highest overall performance for the Area and Boundary metrics is achieved by the VG-HuBERT$_3$ model, while the highest performance for the Word metrics is achieved by the VG-HuBERT$_4$ model. For both VG-HuBERT and VG-W2V2, we find that reinitializing the last few layers before training on the visual grounding task is highly beneficial, which is in line with the results found by~\cite{Pasad2021LayerWiseAO}. A more extensive comparison of layer reinitialization is shown in Table~\ref{tab:diff_reinit}. ResDAVEnet-VQ, W2V2, and HuBERT all have a low A-score, mainly due to a low tIoU, which indicates their relative inability to segment words; this is further confirmed by their low boundary F1 scores. In Figure~\ref{fig:girl_lighthouse}, we visually see that the attention segments produced by HuBERT are highly fragmented and do not clearly correspond to words. Despite this, both HuBERT and W2V2 achieve nontrivial word detection results, 
outperforming the ResDAVEnet-VQ model but significantly underperforming VG-HuBERT, VG-W2V2, and FaST-VGS, all of which learn more than twice as many word detectors as HuBERT. This indicates that the masked language modeling objective used to train HuBERT and W2V2 does result in representations strong enough to capture word identity, but the visual grounding objective appears to be significantly more powerful in this respect.  Additionally, the visual grounding objective leads to the emergence of word segmentation ability, whereas the masked language modeling objective does not. Finally, in comparing to FaST-VGS and FaST-VGS+ we note that these models use a strongly supervised vision module (Faster RCNN), whereas VG-HuBERT and VG-W2V2 are fully self-supervised and yet still outperform FaST-VGS and FaST-VGS+ across all metrics.

In Figures~\ref{fig:diff_layers} and~\ref{fig:diff_k}, we examine the word detection performance of various models across different Transformer layers and as we vary the number of K-means clusters. Although increasing the number of clusters does lead to a larger number of detectors (Figure~\ref{fig:diff_k}), we see significant diminishing returns and thus fix $K=4096$ in all of our experiments, unless stated otherwise. In Figure~\ref{fig:diff_layers}, we see that for all models, word detection performance is best in the middle and upper half of the model, which is also consistent with the analysis of~\cite{Pasad2021LayerWiseAO} Finally, we compare the word detection performance of VG-HuBERT when using oracle word boundaries to determine segments (rather than the model's thresholded self-attention), which we counterintuitively find hurts the model's word detection ability. Combined with the observation that the attention segments tend to concentrate at the nucleus of words, we hypothesize that the model's contextualization is pushing word identity information towards the temporal center of each word.

\begin{table}[ht]
\vspace{-3mm}
\caption{Results on ZeroSpeech 2020 spoken term discovery.}\label{tab:zs20}
\vspace{-6mm}
\begin{center}
\resizebox{\columnwidth}{!}{%
\begin{tabular}{lcccclll}
\toprule
& Words & NED & Cov & M-score & Prec. & Rec. & $F_1$  \\
\midrule
J.V.~\cite{jansen2011efficient} & 18821 &\textbf{32.4} &7.9 &14.1& 32.1 &3.2 &5.9 \\
Granada et al. & 92544 & 72.0 & 76.9 & 41.1 & 30.7 & 43.8 & 36.1 \\
Räsänen et al. & \textbf{321603} & 52.5 & 28.7 & 35.8 & 23.9 & 24.2 & 24.1 \\
ES-KMeans~\cite{kamper17} & 42473 & 73.2 & \textbf{100.0} & 42.3 & \textbf{49.4} & 66.8 & \textbf{56.7} \\
SEA~\cite{Bhati2020SelfExpressingAF} & \underline{240033} & 89.5 & \underline{99.5} & 19.0 & 32.5 & \underline{78.9} & 46.1 \\
PDTW~\cite{Rsnen2020UnsupervisedDO} & 85425 & 48.2 & 85.4 & \underline{64.5} & 26.5 & \textbf{88.2} & 40.8 \\
$\text{VG-HuBERT}_3$ (Ours) & 104696 & \underline{42.5} & 95.4 & \textbf{71.8} & \underline{44.7}& 54.2&\underline{49.0}\\
\bottomrule
\end{tabular}}
\end{center}
\vspace{-5mm}
\end{table}
\begin{table}[ht]
	\caption{Unsupervised word segmentation on Buckeye test set.}
	\vspace{-3mm}
\resizebox{\columnwidth}{!}{%
	\begin{tabular}{lccccc}
		\toprule
		& \multicolumn{4}{c}{{Boundary}} & \multicolumn{1}{c}{Token} \\
		\cmidrule(r){2-5} \cmidrule(l){6-6}
        {Model} & {Prec.}  & {Rec.}  & {$F_1$}  & {$R$-val.} & {$F_1$} \\
		\midrule
		Adaptor gram.~\cite{Jansen2013ASO} & 15.9 & \textbf{57.7} & 25.0 & -139.9 & 4.4 \\
		 SylSeg~\cite{rasanen2015unsupervised} & 27.7 & 28.9 & 28.3 & 37.7 & 19.3 \\
		 ES-KMeans~\cite{kamper17} & 30.3 & 16.6 & 21.4 & 39.1 & 19.2 \\
		 BES-GMM~\cite{besgmm} & 31.5 & 12.4 & 17.8 & 37.2 & 18.6 \\
		 SCPC~\cite{Bhati2021SegmentalCP} & 36.9 & 29.9 & 33.0 & \underline{45.6} & {-} \\
		 mACPC~\cite{Cuervo2021ContrastivePS} &  \textbf{42.1} & 30.3 & 35.1 &  \textbf{47.4} & {-} \\
		DPDP~\cite{Kamper2022WordSO} & 35.3 &  37.7 &  \textbf{36.4} & 44.3 &  \textbf{25.0} \\
      $\text{VG-HuBERT}_3$ (Ours) & 36.2 & 32.2 & 34.1 & \underline{45.6}& \underline{24.1} \\
		\bottomrule
	\end{tabular}}
	\label{tab:wordseg_buckeye}
	\vspace{-6mm}
\end{table}

\textbf{Results on Buckeye and ZeroSpeech 2020.} Finally, we test our best performing model ($\text{VG-HuBERT}_3$) \footnote{we use model's best word segmentation hyperparams, max-pooling instead of mean-pooling for segment features, and 16384 clusters for K-means.}, on two additional benchmarks: the ZeroSpeech 2020 spoken term discovery track (English) and the Buckeye word segmentation task\footnote{we make use of \href{https://github.com/kamperh/vqwordseg}{https://github.com/kamperh/vqwordseg}}.

Table~\ref{tab:zs20} shows model performance on ZeroSpeech, where following the official challenge leaderboard and \cite{Rsnen2020UnsupervisedDO}, we report number of discovered words (Word), phoneme level normalized edit distance (NED) of the discovered segment pairs, coverage of the discovered segments (Cov) w.r.t. all utterances in the given corpus, and also word segmentation performance (precision, recall, F-score). We also report M-score~\cite{Rsnen2020UnsupervisedDO}, which is the harmonic mean of (100 - NED) and Cov that summarizes the trade-off between purity and coverage (similar to A-score we used for previous evaluation). We see that $\text{VG-HuBERT}_3$ achieves the highest M-score and second best boundary $F_1$.

Table~\ref{tab:wordseg_buckeye} shows word segmentation performance on Buckeye test set, where we again report boundary precision, recall, and $F_1$. 
Token $F_1$ is harsher metric where both left and right boundaries of a word need to be successfully predicted. We see that our $\text{VG-HuBERT}_3$ performs competitively compared to the state of the art approaches on Boundary $F_1$, $R$-value, and Token $F_1$. 

Importantly, we note that all other models in Table~\ref{tab:zs20} and ~\ref{tab:wordseg_buckeye} are unimodal models, i.e. they are only trained on the audio modality, while our model uses the visual modality during training (visual modality is not needed during testing).
On the other hand, these other models are explicitly designed for word discovery and segmentation via specific mechanisms or learning objectives, while our model is instead trained with a cross-modal grounding objective. Our experiments demonstrate that this objective, combined with the Transformer architecture pre-trained on a masked language modeling task, facilitates \textit{emergent} word discovery ability from raw speech signals.

\vspace{-2mm}
\section{Conclusion and Future Directions}
\vspace{-2mm}
In this paper, we demonstrated that training the popular HuBERT and wav2vec2.0 models on a visual grounding task resulted in them learning to automatically discover (localize, segment, and identify) words in the speech signal. We verified that this same ability did not exist in the models prior to them undergoing training on the visual grounding task, suggesting that a combination of multiple self-supervised objectives may be essential in order for a computational model to learn the structure of spoken language from untranscriped speech audio.

In our future work, we plan to use the word-level representations learned by our model to anchor the learning of sub-word structure (e.g. automatically learning a pronunciation lexicon in terms of discovered speech units) as well as the higher level syntactic structure of spoken language. We also plan to investigate whether a similar method to the one we propose could facilitate visual object discovery, and also explore applications of our method to low-resource ASR.

\vspace{-2mm}
\section{Acknowledgements}
\vspace{-2mm}
We thank Abdelrahman Mohamed, Karen Livescu, Haoyue Freda Shi, Cheng-I Jeff Lai, and Bowen Shi for their insights.

\bibliographystyle{IEEEtran}
\small
\bibliography{main}

\end{document}